\documentstyle[epsfig]{elsart}
\newcommand{\lag}{{\mathcal{L}}}
\newcommand{\covder}{\!\not \! \!D}
\newcommand{\ltsim}{\lower3pt\hbox{$\, \buildrel < \over \sim \, $}}
\newcommand{\gtsim}{\lower3pt\hbox{$\, \buildrel > \over \sim \, $}}
\newcommand{\glt}{\lower3pt\hbox{$\, \buildrel < \over > \, $}}
\renewcommand{\O}{{\mathcal{O}}}

\begin{document}
\begin{frontmatter}

\rightline{UG--FT--116/00}
\rightline{MIT-CTP-2996}
\rightline{July 2000}  

\title{Effective description of quark mixing}
\author[Granada]{F. del Aguila},
\author[MIT]{M. P\'erez-Victoria} \and
\author[Granada]{J. Santiago}
\address[Granada]{Departamento de F\'{\i}sica Te\'{o}rica y del Cosmos \\
Universidad de Granada \\
E-18071 Granada, Spain}
\address[MIT]{Center for Theoretical Physics \\ Massachusetts
Institute of Technology \\ Cambridge, MA 02139, USA}
\date{\today}
\begin{abstract}
We use the effective Lagrangian formalism to describe 
quark mixing. The 
new $W^\pm$, $Z$ and $H$ couplings generalizing the CKM matrix and the
GIM mechanism fulfil relations and inequalities which allow to 
discriminate among different SM extensions. As a by-product we give a
useful parametrization of the generalized CKM matrix. We also show
that the largest possible departures from the SM predictions 
result from heavy exotic fermions, which can induce, for example, 
top FCNC large enough to be observable at future colliders. 
\end{abstract}
\end{frontmatter}

The mixing of fermions with the same quantum numbers provides a very
sensitive window to new physics. Two distinguished examples are
the prediction of the existence of the charm quark by Glashow,
Iliopoulos and 
Maiani (GIM)~\cite{papiro6} from the absence of flavour changing neutral
currents (FCNC), and the recent results on neutrino oscillations
indicating a non-zero mass for the neutrinos. Here we are interested
in the quark sector. The Standard Model (SM) predicts a definite
pattern of quark mixing: FCNC are absent at tree level and suppressed
at one loop, the mixing in charged currents is given by the unitary
Cabibbo-Kobayashi-Maskawa (CKM) matrix~\cite{papiro5} and the source of
all CP violation is the unique phase in this matrix. The fact
that some of the CKM matrix elements, especially the ones involving
the top, are poorly known at present, implies that the unitarity of
the $3\times 3$ CKM 
matrix has yet to be tested. Moreover, little is known so far about
possible FCNC for the top. Fortunately the situation is expected to
improve in the near future. Many ongoing and future experiments
(Tevatron, B factories, LHC) will be able to test the unitarity
triangle~\cite{triangle} and to measure top couplings to within $\sim
1\%$ of the 
typical size of a weak coupling, thus improving present bounds by more
than one
order of magnitude~\cite{papiro7}. The usual procedure to analyse these
experiments is to assume the
SM and determine the corresponding parameters from the
measurements. New physics would then manifest itself as
inconsistencies arising from the fact that the experiments
overconstrain this 
parametrisation. For instance, the non-closure of the unitarity
triangle would directly indicate non-standard physics. Obviously, this
method is insufficient to learn about the new physics if a new effect
is found. Moreover, a na\"{\i}ve use of the SM ansatz can
be misleading when interpreting results hinting a closed unitarity
triangle. In order to analyse and discuss new
physics it seems more convenient to work with a more general
parametrisation from the very beginning. A convenient one
should reflect which parameters are small and vanish in the
SM limit. Furthermore, one would like to know what relations between
neutral and charged currents can be imposed and what bounds should
be expected on general 
grounds (symmetry, dimensional analysis, etc.). All these
issues can be best dealt with using effective theory
techniques, and this is what we do in this paper. We also argue that
only models with extra vector-like quarks can give large new effects
beyond the SM.

The SM should be understood as the lowest dimension part in the
expansion of an effective Lagrangian that describes any physics
below a certain scale $\Lambda$ (see Ref.~\cite{papiro2} and references
there in). 
This effective Lagrangian can describe a large class of
SM extensions, including models with extra gauge
interactions, new vector bosons, fermions or scalars with
or without supersymmetry, and in four or higher
dimensions~\cite{papiro3,papiro4}. $\Lambda$ is a characteristic
scale of the high energy theory, typically given by 
the lowest thresholds of the non-standard particles.
An effective Lagrangian can describe both decoupling and
non-decoupling non-standard physics (in the second case the effective
description is mandatory). Here we shall assume that the low energy
scalar sector is given by an elementary (not too heavy) Higgs and
therefore work in a decoupling scenario\footnote{The same analysis of this
paper could be repeated starting with the chiral
SM~\cite{papiro8}. The results 
should be analogous since the flavour structure is not
altered.}. Moreover, it is sufficient to consider only the SM Higgs
for the following reasons: Scalar singlets which can acquire a vacuum
expectation value (vev) do
not transform under the SM and are then flavour blind. Other
multiplets like  
triplets can only get very small vevs $v_t$ (due to the constraints on
the $\rho$ parameter) and their effects are
suppressed by powers of $\frac{v_t}{v}$, where $v$ is the SM
vev. Finally, for any number of 
Higgs doublets we identify the minimal SM one with the combination
getting a vev. 
 
The experimental precision and the scale $\Lambda$ determine the order
to be considered in the expansion
\begin{equation}
\lag^{\mathit{eff}}=\lag_4+\frac{1}{\Lambda}\lag_5 +
\frac{1}{\Lambda^2} \lag_6+\ldots\quad .\label{lag:456}
\end{equation}
$\lag_4$ is the SM Lagrangian and $\lag_{5,6,\ldots}$ contain
operators ${\mathcal{O}}_x$ of dimension 5, 6,$\ldots$,
respectively. All these operators are invariant under the gauge group
$SU(3)_C\times SU(2)_L \times U(1)_Y$. 
There is an extensive literature on the operators in $\lag_5$ and
$\lag_6$ 
and on their experimental constraints. In Ref.~\cite{papiro3} it is
shown that, assuming lepton and
baryon number conservation, no dimension 5 operator can be constructed
with the SM fields, while there are 81 independent gauge invariant
operators of dimension 6 (up to flavour indices) included in
\begin{equation}
\lag_6=\alpha_x {\mathcal{O}}_x+\mathit{h.c.} \, .  
\label{lag:6}
\end{equation}
The coefficients $\alpha_x$ parametrise the new physics beyond the SM. 
Other operators are allowed, but can be transformed into the ones
in~\cite{papiro3} by using the equations of motion of $\lag_4$. This
does not change S matrix elements to order $1/\Lambda^2$~\cite{papiro4}.
It should be noted that the Lagrangian $\lag^{\mathit{eff}}$ can be
used beyond the classical 
level. Quantum corrections can be computed systematically as
$\lag^{\mathit{eff}}$ is renormalizable in the modern sense that there
is a counterterm available to cancel every
infinity~\cite{weinberg}. A good and consistent approximation is to
consider the full one-loop 
corrections to $\lag_4$, but work at tree level whenever an operator in
$\lag_6$ is inserted. The effect of the SM radiative corrections
should be taken into account when applying our results to precision
data. 

In order to compare with
experiment we have to consider the spontaneous symmetry
breaking (SSB) of the electroweak gauge symmetry, which introduces a
new dimensionful parameter: the electroweak vev $v\sim 250\;
\mbox{GeV}$. Then $\lag_6$ gives
contributions  $v^2 \lag_4^\prime$, 
$v \lag_5^\prime$ and $\lag_6^\prime$, where 
$\lag_d^\prime$ contains 
operators of dimension $d$ invariant
under the unbroken $SU(3)_C\times U(1)_Q$ gauge symmetry. Quark mixing
can occur in two-fermion and four-fermion operators. The latter have
dimension 6 and the former (after SSB), dimension 4, 5 and
6. Operators with the same fields but with a different dimension
after SSB can in principle be  
distinguished experimentally, since they lead to a different momentum
dependence. Four-fermion operators can give
non-standard mixing in kaon or B meson experiments, quark-pair
production, etc. They are generated in many extensions of the SM. 
Here, however, we shall concentrate on the trilinear couplings
$V\bar{q}q^\prime, \; V=Z,W^\pm$, and 
$H\bar{q}q^\prime$, which generalize the CKM matrix
and the diagonal couplings in neutral currents. 
These vertices can be measured independently and
may have a different origin than the four-fermion
operators.

The operators $\O_x$ are generated by virtual exchange of the heavy
modes in the high-energy theory.
It is useful to distinguish the operators which are generated at
tree level from those which are generated only by loop
diagrams~\cite{papiro4}. The latter are suppressed by powers of
$1/16 \pi^2$, at least when the heavy theory is weakly
interacting. We can then consider only operators generated at tree
level, since we are mainly interested in the largest SM deviations,
which might be observed within  
the precision of future experiments. This restriction
reduces the list of operators relevant to trilinear couplings to the
seven operators collected in Table~\ref{operators}. These
operators give only, after SSB, dimension 4 (i.e., $\sim
v^2/\Lambda^2$) operators and operators of dimension 5 of the form
$\partial H\,\bar{q} q^\prime$. Hence, magnetic-moment type operators
and other operators 
with extra momentum dependence are not generated at tree level.
On the other hand, it is important to observe that these seven
operators are the only ones contributing to the quark sector of
$\lag^\prime_4$~\cite{papiro3}. Therefore, even though we are mainly
concerned with large effects in the top couplings, all the results
in this paper for the couplings $V\bar{q}q^\prime$, $H\bar{q}q^\prime$
apply to any kind of new physics, independently of whether it
contributes at tree level or not. There are also other operators which
would 
redefine the $Z$, $W^\pm$ and $H$ fields in the trilinear quark
couplings, 
but they are flavour blind and need not be taken into account for
our purposes. No flavour change occurs in the dimension 4
couplings to the photon and gluons due to the exact $U(1)_Q$ and
$SU(3)_C$ symmetry. 
\begin{table*}[!h]
\caption{Dimension 6 operators correcting trilinear
$V\bar{q}q^\prime$, $V=W^\pm,Z$, and $H\bar{q}q^\prime$
vertices. (See Ref.~\cite{papiro3} for notation.)}
\label{operators}\vspace{0.5cm}
\begin{tabular}{ll}
$({\mathcal{O}}_{\phi q}^{(1)})^{ij}= 
(\phi^\dagger i
D_\mu\phi)(\bar{q}^i_L\gamma^\mu q^j_L)$
\\
$({\mathcal{O}}_{\phi q}^{(3)})^{ij}= 
(\phi^\dagger \tau^I i
D_\mu\phi)(\bar{q}^i_L\gamma^\mu \tau^I q^j_L)$
\\
 $({\mathcal{O}}_{\phi u})^{ij}= (\phi^\dagger i
D_\mu\phi)(\bar{u}^i_R\gamma^\mu u^j_R)$
\\
 $({\mathcal{O}}_{\phi d})^{ij}= (\phi^\dagger i
D_\mu\phi)(\bar{d}^i_R\gamma^\mu d^j_R)$
\\
$({\mathcal{O}}_{\phi \phi})^{ij}= (\phi^T \epsilon i
D_\mu\phi)(\bar{u}^i_R\gamma^\mu d^j_R)$
\\
$({\mathcal{O}}_{u \phi})^{ij}= 
(\phi^\dagger \phi)(\bar{q}^i_L \tilde{\phi}u^j_R)$
\\
$({\mathcal{O}}_{d \phi})^{ij}= 
(\phi^\dagger \phi)(\bar{q}^i_L  \phi d^j_R)$
\end{tabular}
\vspace{0.5cm}
\end{table*}

In the following we extend the analysis of Ref.~\cite{papiro3} to
describe quark mixing.
We generalize the CKM matrix and the GIM
mechanism, and  find relations and bounds fulfilled by the trilinear
couplings. As an illustration of how the general description can be
employed to study particular models, we discuss a simple
extension of the SM with an extra exotic up quark isosinglet
$T$~\cite{papiro10,papiro9}. This example also shows
that a top mixing large enough to be observable at future
colliders can actually be produced in explicit models~\cite{papiro11}.
Finally, we argue that the largest possible departure from the SM is
obtained in models with heavy vector-like fermions. These models are
analised in a subsequent paper~\cite{papiro9}.

The seven operators in Table~\ref{operators}, together with
Eqs. (\ref{lag:456}) and (\ref{lag:6}), describe the large
contributions of an arbitrary SM extension to the 
trilinear quark couplings, and any contribution (large or small) to
non-derivative trilinear quark couplings. Note that the actual value
of the coefficients $\alpha_x$ depends on the specific basis one
uses for the quark fields, which is not completely fixed by the
requirement of canonical kinetic terms and diagonal gauge terms before
SSB. The physical results are independent of the choice of basis, but
for definiteness we shall use here the basis in which the SM Yukawa
couplings of the down quarks are diagonal, real and positive and the
SM Yukawa couplings of the up quarks are of the form
$\lambda^u_{ij}=V_{ij}^\dagger 
\lambda^u_j$ with $\lambda^u_i$ real and positive and $V$ unitary (and
identical to the CKM matrix in the 
SM). After SSB, the quark mass matrices to order $v^2/\Lambda^2$ can
be made diagonal, real and positive by biunitary transformations
$u_{L,R} = U^u_{L,R} \, u^{\prime}_{L,R}$, $d_{L,R} =
U^d_{L,R} \, d^{\prime}_{L,R}$, where the prime is used for mass
eigenstates. We omit it in the following, for all the subsequent
expressions are written in the mass eigenstate basis. Up to the
freedom to redefine the phases of the 
quark mass eigenstates, the diagonalizing matrices read (for
non-degenerate masses, as in the case of the SM)
\begin{equation}
\begin{array}{ll}
U^u_L= V^\dagger ({\mathbf{1}}+ \frac{v^2}{\Lambda^2} A^u_L)\, , ~ & 
U^u_R= {\mathbf{1}} + \frac{v^2}{\Lambda^2} A^u_R \, , \\
U^d_L= {\mathbf{1}} + \frac{v^2}{\Lambda^2} A^d_L \, ,  & 
U^d_R= {\mathbf{1}} + \frac{v^2}{\Lambda^2} A^d_R \, ,
\end{array}
\end{equation}
where ${\mathbf{1}}$ is the identity matrix and
$A^{u,d}_{L,R}$ are antihermitian matrices given by
\begin{eqnarray}
(A^u_L)_{ij} & = & \frac{1}{2}(1-\frac{1}{2} \delta_{ij})
\frac{\lambda^u_i (V 
\alpha_{u\phi})^\dagger_{ij} + (-1)^{\delta_{ij}}
(V\alpha_{u\phi})_{ij} \lambda^u_j} 
{(\lambda^u_i)^2-(-1)^{\delta_{ij}}(\lambda^u_j)^2} \, ,\nonumber \\
(A^u_R)_{ij} & = & \frac{1}{2}(1- \frac{1}{2} \delta_{ij})
\frac{\lambda^u_i (V 
\alpha_{u\phi})_{ij} + (-1)^{\delta_{ij}}
(V\alpha_{u\phi})^\dagger_{ij} \lambda^u_j} 
{(\lambda^u_i)^2-(-1)^{\delta_{ij}}(\lambda^u_j)^2} \, ,  \\
(A^d_L)_{ij} & = & \frac{1}{2}(1- \frac{1}{2} \delta_{ij})
\frac{\lambda^d_i 
(\alpha_{d\phi})^\dagger_{ij} + (-1)^{\delta_{ij}}
(\alpha_{d\phi})_{ij} \lambda^d_j} 
{(\lambda^d_i)^2-(-1)^{\delta_{ij}}(\lambda^d_j)^2} \, ,\nonumber \\
(A^d_R)_{ij} & = & \frac{1}{2}(1- \frac{1}{2} \delta_{ij})
\frac{\lambda^d_i 
(\alpha_{d\phi})_{ij} + (-1)^{\delta_{ij}}
(\alpha_{d\phi})^\dagger_{ij} \lambda^d_j}  
{(\lambda^d_i)^2-(-1)^{\delta_{ij}}(\lambda^d_j)^2} \, . \nonumber
\end{eqnarray}
Our initial choice of basis implies that only $U^u_L$
is non-trivial at order 1. $\alpha_{u\phi}$ and $\alpha_{d\phi}$ are
the coefficients of the 
operators that contribute to the quark masses (see Table~1). 
In terms of mass eigenstates, the generic quark couplings have the
following form: 
\begin{eqnarray}
\lag^Z&=&-\frac{g}{2\cos \theta_W} \left( 
\bar{u}^i_L X^{uL}_{ij}
\gamma^\mu u^j_L+
\bar{u}^i_R X^{uR}_{ij}
\gamma^\mu u^j_R \right.\nonumber \\
&&\left.-\bar{d}^i_L X^{dL}_{ij}
\gamma^\mu d^j_L-
\bar{d}^i_R X^{dR}_{ij}
\gamma^\mu d^j_R - 2\sin^2\theta_W J^\mu_{\mathrm{EM}}\right) Z_\mu,
\nonumber \\ 
\lag^W&=&-\frac{g}{\sqrt{2}}(
\bar{u}^i_L W^{L}_{ij}
\gamma^\mu d^j_L+
\bar{u}^i_R W^{R}_{ij}
\gamma^\mu d^j_R)W^+_\mu +\textit{h.c.}
, \label{lag:ZWH}\\
\lag^H&=&-\frac{1}{\sqrt{2}}(
\bar{u}^i_L Y^{u}_{ij}
 u^j_R+
\bar{d}^i_L Y^{d}_{ij} d^j_R + \mathit{h.c.} ) H
\nonumber \\ 
&&+ \left( 
\bar{u}^i_L Z^{uL}_{ij}
\gamma^\mu u^j_L+
\bar{u}^i_R Z^{uR}_{ij}
\gamma^\mu u^j_R-\bar{d}^i_L Z^{dL}_{ij}
\gamma^\mu d^j_L-
\bar{d}^i_R Z^{dR}_{ij}
\gamma^\mu d^j_R \right) i\partial_\mu H. \nonumber 
\end{eqnarray}
The unbroken $U(1)_Q$ protects the terms proportional to
$J^\mu_{\mathrm{EM}}$. 
The expressions to order $1/\Lambda^2$ of the coupling matrices 
$X$, $W$, $Y$ and $Z$ in terms of the coefficients $\alpha_x$ are:
\begin{eqnarray}
X^{uL}_{ij}&=& \delta_{ij}-\frac{1}{2}\frac{v^2}{\Lambda^2}
V_{ik}(\alpha^{(1)}_{\phi q}+\alpha^{(1)\dagger}_{\phi q}
-\alpha^{(3)}_{\phi q}-\alpha^{(3)\dagger}_{\phi q}
)_{kl}V^\dagger_{lj}, \nonumber \\
X^{uR}_{ij}&=& -\frac{1}{2}\frac{v^2}{\Lambda^2}
(\alpha_{\phi u}+\alpha_{\phi u}^\dagger
)_{ij}, \nonumber \\
X^{dL}_{ij}&=& \delta_{ij}+\frac{1}{2}\frac{v^2}{\Lambda^2}
(\alpha^{(1)}_{\phi q}+\alpha^{(1)\dagger}_{\phi q}+
\alpha^{(3)}_{\phi q}+  
\alpha^{(3)\dagger}_{\phi
q})_{ij}, \nonumber \\
X^{dR}_{ij}&=& \frac{1}{2}\frac{v^2}{\Lambda^2}
(\alpha_{\phi d}+\alpha_{\phi d}^\dagger
)_{ij},  \label{couplings}\\
W^{L}_{ij}&=& \tilde{V}_{ik}\left(\delta_{kj}+\frac{v^2}{\Lambda^2}
(\alpha^{(3)}_{\phi q})_{kj}\right), \nonumber \\
W^{R}_{ij}&=& -\frac{1}{2}\frac{v^2}{\Lambda^2}
(\alpha_{\phi \phi})_{ij}, \nonumber \\
Y^{u}_{ij}&=& \delta_{ij}\lambda^u_j - \frac{v^2}{\Lambda^2} 
\left(V_{ik}(\alpha_{u \phi})_{kj} +\frac{1}{4} \delta_{ij}
[V_{ik} (\alpha_{u \phi})_{kj} + (\alpha_{u \phi}^\dagger)_{ik}
V^\dagger_{kj}]\right), 
\nonumber \\ 
Y^{d}_{ij}&=& \delta_{ij}\lambda^d_j - \frac{v^2}{\Lambda^2} 
\left((\alpha_{d \phi})_{ij}  + \frac{1}{4} \delta_{ij} (\alpha_{d
\phi}+\alpha_{d\phi}^\dagger)_{ij} \right) , \nonumber \\ 
Z^{uL}_{ij}&=& -\frac{1}{2}\frac{v}{\Lambda^2}
V_{ik}(\alpha^{(1)}_{\phi q}-\alpha^{(1)\dagger}_{\phi q}
-\alpha^{(3)}_{\phi q}+\alpha^{(3)\dagger}_{\phi q}
)_{kl}V^\dagger_{lj}, \nonumber \\
Z^{uR}_{ij}&=& -\frac{1}{2}\frac{v}{\Lambda^2}
(\alpha_{\phi u}-\alpha_{\phi u}^\dagger
)_{ij}, \nonumber \\
Z^{dL}_{ij}&=& \frac{1}{2}\frac{v}{\Lambda^2}
(\alpha^{(1)}_{\phi q}-\alpha^{(1)\dagger}_{\phi q}+
\alpha^{(3)}_{\phi q}-  
\alpha^{(3)\dagger}_{\phi
q})_{ij}, \nonumber \\
Z^{dR}_{ij}&=& \frac{1}{2}\frac{v}{\Lambda^2}
(\alpha_{\phi d}-\alpha_{\phi d}^\dagger
)_{ij}.\nonumber
\end{eqnarray}
We have introduced the unitary matrix
\begin{equation}
\tilde{V} = V+ \frac{v^2}{\Lambda^2} (V A_L^d - A_L^u V) \, .
\end{equation}
Note that, to order $1/\Lambda^2$, we can substitute $V$ by
$\tilde{V}$ everywhere in Eq.~(\ref{couplings}), so that the different
couplings 
depend on only one unitary matrix.
These couplings incorporate features that are
forbidden in the SM, namely, FCNC,
right-handed neutral currents not proportional to $J^\mu_{\mathrm{EM}}$,
right-handed charged currents, and left-handed charged currents which
are not described by a unitary matrix. These effects are allowed in
general to order $1/\Lambda^2$. We stress that these trilinear
couplings can be
directly determined from processes involving the $\bar{q}
q^{\prime} V$ and $\bar{q} q^{\prime} H$ vertices in which the final
$V$ or $H$ is 
observed. For the top 
quark this will be possible in large hadron colliders. 
Although the trilinear couplings also contribute to four-fermion
processes (such as mixing of neutral mesons), one should remember that
four-fermion operators may contribute in this case as
well~\cite{papiro10,BSW}. Nevertheless, one can still use these 
processes to put limits on the trilinear couplings under the
assumption that no strong cancellations between cubic and quartic
couplings take place. 

The CKM matrix is now an arbitrary $3\times 3$ matrix and can be
written as the product of a unitary matrix $V^L$ times a hermitian
matrix $H^L$,
\begin{equation}
W^L=V^L H^L \, . \label{WL:HLVL}
\end{equation}
Alternatively one can write $W^L={H^L}^\prime V^L$, with
${H^L}^\prime = V^L H^L V^{L \dagger}$. Basically, the decomposition
in Eq.~(\ref{WL:HLVL}) is better suited to study the down sector,
whereas the other one is more convenient for the up sector.
For definiteness, we use the first decomposition in the
following discussion.
After using all the freedom to rotate the quark phases, $W^L$ depends
on 13 parameters. It is possible to express $V^L$ in any of the existing
parametrizations of the CKM matrix. $V^L$ then depends on 3 angles and
1 phase and $H^L$ depends on 6 real numbers and
3 phases.
Since the SM deviations are small perturbations, $H^L$ can be expanded
around the identity. In our effective Lagrangian description,
\begin{equation}
H^L={\mathbf{1}}+\frac{v^2}{2\Lambda^2} e^{-i\theta^d}
\left(\alpha_{\phi q}^{(3)} +  
\alpha_{\phi q}^{(3) \, \dagger}\right) e^{i \theta^d} \label{hache}
\, ,
\end{equation}
and $V^L= e^{-i\theta^u} \tilde{V} ( {\mathbf{1}} +
\frac{v^2}{2\Lambda^2} (\alpha_{\phi q}^{(3)} - \alpha_{\phi q}^{(3)
\, \dagger} ) )\; e^{i\theta^d}$ \, ,  
where $e^{i\theta^{u,d}}$ are the diagonal phase matrices that bring
$V^L$ to the desired form. On the other hand, $V^L$ can be expanded in
the Cabibbo angle $\lambda$
\textit{\`{a} la} Wolfenstein~\cite{papiro12}\footnote{It should be
noted that the Wolfenstein parametrization introduces an artificial
loss of unitarity which should not be confused with the physical loss
of unitarity contained in $H^L$. This effect can be made arbitrarily
small by considering 
sufficiently high orders in $\lambda$. For many purposes, like CP
violation in neutral kaon decays or high precision CP studies of
neutral B-meson decays, one should incorporate corrections
proportional to $\lambda^4$ and $\lambda^5$.}.
To order $\lambda^3$, the generalized CKM matrix reads 
\begin{equation}
W^L=
\left(\begin{array}{ccc}
1-\frac{\lambda^2}{2} & \lambda & A \lambda^3 (\rho-i\eta) \\
-\lambda & 1-\frac{\lambda^2}{2} & A \lambda^2 \\
A \lambda^3 (1-\rho-i\eta) & -A\lambda^2 & 1
\end{array}\right)
\left(\begin{array}{ccc}
1 + \omega_{11} & \omega_{12} & \omega_{13} \\
\omega_{12}^* & 1 + \omega_{22} & \omega_{23} \\
\omega_{13}^* & \omega_{23}^* & 1 + \omega_{33} \end{array} \right) \,
, \label{W:parametrisation}
\end{equation}
where the diagonal elements of $H^L$, $\omega_{ii}$, are small real
numbers and the off-diagonal ones, $\omega_{ij}$ for $i\not = j$, are
small complex numbers.
The parameters $\omega_{ij}$ vanish in the SM
(they are $\sim v^2/\Lambda^2$). The $\omega_{ii}$ and the
$|\omega_{ij}|$, $i\not = j$, are invariant under
redefinitions of the quark phases, and depend on the hermitian
part of the coefficient of a single operator, as shown
in Eq.~(\ref{hache}). 
The unitarity relations are modified at order $\sim v^2/\Lambda^2$. The
non-closure of the unitarity triangles is directly given by the
off-diagonal $\omega_{ij}$. For example, we have
\begin{equation}
W_{ud} W_{ub}^* + W_{cd} W_{cb}^* + W_{td} W_{tb}^* = 2 \omega_{13}^*
\, . 
\end{equation}
The remaining $\omega_{ij}$ can be determined, in principle,
from measurements of the other unitarity relations. Observe that the
different ``triangles'' are not equivalent for a non-unitary CKM
matrix. On the other hand, there are now four independent physical
phases that can induce CP violation. In the alternative decomposition
$W^L={H^L}^\prime V^L$, ${H^L}^\prime$ is analogously parametrised by 
$\omega^\prime_{ij}=V^L_{ik} \omega_{kl} {V^L}^\dagger_{lj}$, where 
$\omega^{(\prime)}_{ij}={\omega^{(\prime)}}^*_{ji}$ for $i>j$. Note
that $\omega^\prime_{ij}=\omega_{ij}$ up to terms suppressed by
the Cabibbo angle. Care should be taken in the extraction of the
effective couplings $W^L_{ij}$ from the data, as it may be affected by
the presence of right-handed charged currents.

The mixing matrix for right-handed charged currents can be similarly
decomposed as $W^R= V^R H^R$, where $H^R$ is of order
$v^2/\Lambda^2$. Note, however, that after bringing $V^L$ 
to a given form there is no freedom to redefine $V^R$
while keeping the masses real and positive. Therefore $W^R$ has 18
independent real (and small) parameters.

The couplings to the massive vector bosons, $X_{ij}$ and $W_{ij}$ in
Eq. (\ref{couplings}),
are well-known for the 5 lightest quarks and will be precisely
measured at future colliders for the top quark.  
A relevant question  is what can be
said about the SM deviations on general grounds. There are 
6 types of couplings $X_{ij}$, $W_{ij}$ 
in Eq. (\ref{couplings}) and they are a function of 5 types of
operators and a unitary matrix. Combining the different expressions we
find the general relations
\begin{eqnarray}
X^{uL}_{ij}&=& 2 W^L_{ik}W^{L\dagger}_{kj}- W^L_{ik}X^{dL}_{kl}
W^{L\dagger}_{lj}, \label{XWW}\\ 
X^{dL}_{ij}&=& 2 W^{L\dagger}_{ik}W^{L}_{kj}- W^{L\dagger}_{ik}X^{uL}_{kl}
W^{L}_{lj}. 
\end{eqnarray}
Other relations may be fulfilled by the couplings to the $Z$ and $W^{\pm}$
for particular classes of models. A possibility is that the new
physics is such that
\begin{equation}
\alpha^{(1)}_{\phi q}=a\alpha^{(3)}_{\phi q},\label{prop}
\end{equation}
with $a$ a number. This occurs
whenever all the heavy fields contributing at tree level have the same
statistics and 
transform in the same representation of the gauge group. Then $a$ is
given by Clebsh-Gordan coefficients. This is the case of models with
heavy exotic quarks of just one type~\cite{papiro9}.
If Eq.~(\ref{prop}) holds then there are two additional relations:
\begin{eqnarray}
X^{uL}_{ij}&=& 2 \frac{1+a}{3+a} \delta_{ij}+\frac{1-a}{3+a} W^L_{ik}
X^{dL}_{kl}
W^{L\dagger}_{lj}, \label{XWW:prop:a}\\ 
X^{dL}_{ij}&=& 2 \frac{1-a}{3-a} \delta_{ij}+\frac{1+a}{3-a}
W^{L\dagger}_{ik} 
X^{uL}_{kl}
W^{L}_{lj}, 
\end{eqnarray}
which substituted in Eq. (\ref{XWW}) give
\begin{eqnarray}
X^{uL}_{ij}&=& \frac{1+a}{2} \delta_{ij}+\frac{1-a}{2} W^L_{ik}
W^{L\dagger}_{kj}, \label{XWW:prop:b}\\ 
X^{dL}_{ij}&=&\frac{1-a}{2} \delta_{ij}+\frac{1+a}{2}
W^{L\dagger}_{ik} 
W^{L}_{kj}.  \label{XWW:prop:b2}
\end{eqnarray}
These relations become particularly simple for $a=\pm 1$. In terms of
the parametrisation of $W^L$ in Eqs.~(\ref{WL:HLVL})
and~(\ref{W:parametrisation}), they read
\begin{eqnarray}
X^{uL}_{ij}&=& \delta_{ij} + (1-a) \omega^\prime_{ij}\, , \label{XWu}\\ 
X^{dL}_{ij}&=& \delta_{ij} + (1+a) \omega_{ij} \, . \label{XWd}
\end{eqnarray}
The severe constraints on the couplings of the first five flavours to
the $Z$ imply that $\omega_{ij}$ 
are very small unless $a=-1$, especially for $i \not = j$ (and
$\omega^\prime_{ij}$ for $i,j\not = 3$ are very small unless
$a=1$)~\cite{papiro10,papiro11,papiro1}\footnote{Here and in the
limits in Eq.~(\ref{exp:bounds}) below 
we are using the values in~\cite{papiro1} for neutral currents of
the quarks of the first family. These values give directly the
corresponding couplings to the $Z$ assuming that there are no 
significant contributions of non-standard four-fermion operators to
atomic parity violation experiments.}.
Indeed, we have
\begin{equation}
\begin{array}{lll}
|\omega_{12}| \ltsim 4.1 \times 10^{-5} \, , \hspace{.5cm} &
|\omega_{13}| \ltsim 1.1 \times 10^{-3} \, , \hspace{.5cm} &
|\omega_{23}| \ltsim 1.9 \times 10^{-3} \, , \\
|\omega^\prime_{12}| \ltsim 1.2 \times 10^{-3} \, .
\end{array}
\end{equation} 
Since for each $i,j$,
$\omega^\prime_{ij}$ and $\omega_{ij}$ are of the same order of
magnitude, we can conclude that for models satisfying
Eq.~(\ref{prop}) and $a \not = -1$ 
the top flavour changing couplings $X^{uL}_{it}$, $i\not = t$ must be
very small too, and will not be observed in the next generation of
accelerators. 
Another possibility that leads to simple relations is
\begin{equation}
\alpha^{(3)}_{\phi q}=0 \, . \label{alpha3:0}
\end{equation}
Then we have
\begin{eqnarray}
W^L_{ij} & = & \tilde{V}_{ij}, \label{WX:alpha3:0} \\
\quad X^{uL}_{ij} & = & 2 \delta_{ij}- W^{L}_{ik}
X^{dL}_{kl}W^{L\dagger}_{lj} \, .
\end{eqnarray}

On the other hand, the couplings to the Higgs boson are completely
arbitrary. In the SM there are no derivative couplings of two quarks
and the Higgs and the non-derivative couplings are diagonal and
proportional to the masses of the quarks. At order $1/\Lambda^2$,
however, FCNC mediated by the Higgs are allowed at tree level, and
(non-diagonal) derivative couplings to the Higgs may exist. These
derivative couplings have the same form as the corrections
to the couplings to the $Z$, but they involve the
antihermitian part of the coefficients $\alpha_x$, rather than the
hermitian one. Hence such couplings cannot be generated in models
that only give hermitian contributions.

Besides these relations, the
couplings to the  $Z$, $W^\pm$ and $H$ also satisfy generic
inequalities.
Let us first discuss the bounds on the couplings of quarks to the
$Z$. The coupling matrices $X_{ij}$ are always hermitian, so they can
be diagonalized by a unitary matrix. Let $x_{\mathrm{max}}$ and
$x_{\mathrm{min}}$ be the maximum and the minimum of the eigenvalues
of any $X$. These eigenvalues, and hence $x_{\mathrm{max}}$ and
$x_{\mathrm{min}}$, are
determined by the coefficients 
$\alpha_x$ entering the corresponding expressions in
Eq.~(\ref{couplings}), and are close to 1 (0) for $X^L$
and ($X^R$). Since $X_{ij}-x_{\mathrm{min}} \delta_{ij}$
and $x_{\mathrm{max}} \delta_{ij} -X_{ij}$ are positive semidefinite, 
the following positivity constraints are fulfilled:
\begin{eqnarray}
|X_{ij}-x_{\mathrm{min}} \delta_{ij}|^2 & \leq &
(X_{ii}-x_{\mathrm{min}}) (X_{jj}-x_{\mathrm{min}})\, , \label{X-d} \\
|X_{ij}-x_{\mathrm{max}} \delta_{ij}|^2 & \leq &
(x_{\mathrm{max}}-X_{ii}) (x_{\mathrm{max}}-X_{jj})\, . \label{X-dmax}
\end{eqnarray}
If all the eigenvalues of a certain $X^L$ are $\geq 1$ ($\leq
1$), we can change $x_{\mathrm{min}}$ ($x_{\mathrm{max}}$)
in Eq. (\ref{X-d}) (Eq.~(\ref{X-dmax})) by 1 and obtain an interesting
bound which is
independent of $\Lambda$ and of the details of the $\alpha_x$: 
\begin{equation}
|X^L_{ij}|^2\leq (X^L_{ii}-1)
 (X^L_{jj}-1), ~ \mbox{for $i\not = j$} \, . \label{X-1}  
\end{equation}
Moreover, each diagonal element then fulfils,
\begin{equation}
X^L_{ii} \geq 1 ~(\leq 1) \label{X-1diag}. 
\end{equation}
Correspondingly, if all the
eigenvalues of $X^R$ are positive (negative) semidefinite, we obtain
\begin{eqnarray}
|X^R_{ij}|^2 & \leq & X^R_{ii} X^R_{jj} \, , \label{Xr}
\\
X^R_{ii} & \geq 0 & ~(\leq 0). \label{Xr2}
\end{eqnarray}
Equation~(\ref{Xr}) is trivially satisfied by $X^L$ but in that case it
gives no phenomenologically interesting information. Eqs.~(\ref{X-1})
and~(\ref{Xr}), on 
the other hand, provide stringent constraints. Indeed, 
inserting the atomic parity violation and LEP data~\cite{papiro1} for
$X^{uL,R}_{uu,cc},X^{dL,R}_{dd,bb}$, in Eqs.
(\ref{X-1}) and~(\ref{Xr}) we find the bounds
\begin{equation}    
\begin{array}{cc}
|X^{uL}_{ut}|\leq 0.28, & |X^{uR}_{ut}|\leq 0.14, \\
|X^{uL}_{ct}|\leq 0.14, & |X^{uR}_{ct}|\leq 0.16, 
\end{array}\label{exp:bounds}
\end{equation}
at $90 \%$
C.L.~(we have used the analogous analysis of \cite{papiro11}). 
These limits improve the present production
bounds~\cite{papiro7}. They must be fulfilled, for instance, in
theories with one type of heavy exotic quark~\cite{papiro9,papiro11}.
Even more stringent limits can be obtained from 
charged current data for models fulfilling Eq.~(\ref{prop}). For
example, if $a=-1$ and Eqs.~(\ref{X-1}) and~(\ref{X-1diag}) with
``$\leq$'' hold,   
we can use the measured values of the charged current couplings
$W^L_{ij}$~\cite{papiro11,papiro1} (assuming the absence of
right-handed charged currents) and Eq. (\ref{XWW:prop:b}) to obtain
\begin{equation}
\begin{array}{c}
|X^{uL}_{ut}|\leq 0.05 \, ,\\
|X^{uL}_{ct}|\leq 0.08 \, .  
\end{array} \label{bounds2}
\end{equation}
If the actual values are close to the maximum allowed by the bounds,
the corresponding couplings will be
observed at future colliders~\cite{papiro7}. 
Such values can be reached in explicit models, as the one considered
below.

Similar bounds can be obtained for the charged currents. The
hermitian matrices $W^\dagger W = H^2$
are diagonalized by unitary transformations, which leads to  
inequalities for ${W^{L,R}}^\dagger W^{L,R}$
depending only on the eigenvalues of $H^{L,R}$. (The inequalities for
$W^{L,R} {W^{L,R}}^\dagger$ are identical since both
products give rise to equivalent matrices.)
The equivalent of Eqs.~(\ref{X-d})
and~(\ref{X-dmax}) are obtained  changing $X$ by
$W^\dagger W$ and $x_{\mathrm{max,min}}$ by
$h^2_{\mathrm{max,min}}$, where the latter are the maximum and
minimum of $H^2$. 
If all the eigenvalues of $H^L$
are $\geq 1$ ($\leq 1$) we also have
\begin{eqnarray}
|\sum_k {W^L_{ki}}^* W^L_{kj}|^2 & \leq &  (\sum_k
|W^L_{ki}|^2-1) (\sum_k |W^L_{kj}|^2 -1 )\,,~\mbox{for $i\not = j$}\, 
, \label{WWgt11} \\ 
\sum_k |W^L_{ki}|^2 & \geq & 1 ~(\leq 1)\, , \label{WWgt1}
\end{eqnarray} 
where for clarity we have explicitly indicated the sum over $k$.
Observe that the above condition is satisfied if and only if
$\alpha_{\phi q}^{(3)} + {\alpha_{\phi q}^{(3)}}^\dagger$ is positive
(negative) semidefinite, and this depends only on general features
of the high-energy theory. We show in~\cite{papiro9} that
extensions of the SM with heavy vector-like isotriplets (isosinglets)
satisfy Eq.~(\ref{WWgt1}) with ``$\geq$'' (``$\leq$''). 
The constraints on $W^R$ are trivially satisfied because the matrix
${W^R}^\dagger W^R$ is always positive semidefinite.
Bounds for the couplings to the Higgs can be obtained in a similar
way, but we do not write them here.

The general analysis we have carried out can be applied to particular
models, just by integrating the heavy modes out. As an example we
consider now 
the addition of a heavy vector-like quark isosinglet of charge
$\frac{2}{3}$, $T$.
The full Lagrangian reads
\begin{eqnarray}
\lag&=&\lag^{SM}+\lag_h+\lag_{lh},\label{full:lag:T} 
\nonumber \\
\lag_l^{SM}&=&\bar{q}^i_L i \covder q^i_L+\bar{u}^i_R i \covder
u^i_R+\bar{d}^i_R i \covder d^i_R - (V^\dagger_{ij}\lambda^u_j
\bar{q}^i_L \tilde{\phi}\; u^j_R 
+ \lambda^d_i \bar{q}^i_L \phi\; d^i_R + \mbox{h.c.}) + \ldots , \\
\lag_h&=&\bar{T}_L i\covder T_L+\bar{T}_R i\covder T_R
- M (\bar{T}_L T_R+\bar{T}_R T_L), \nonumber \\
\lag_{lh}&=&-\lambda^\prime_j V_{ji} \bar{T}_R \tilde{\phi}^\dagger
q^i_L 
+ \mathit{h.c.} \, , \nonumber
\end{eqnarray}
where the dots stand for terms not involving the quarks. In this case,
$\Lambda=M$, the mass of the exotic quark. The
integration of the field $T$ gives (see Ref.~\cite{papiro9} for more
details) 
\begin{eqnarray}
\lag_4&=&\lag^{SM}, \label{lag:eff:T}\\
\frac{1}{\Lambda^2 }\lag_6 &=& V^\dagger_{ik} \lambda^{\prime *}_k
\lambda^\prime_l V_{lj} 
\, \bar{q}^i_L \tilde{\phi} \frac{i\covder}{M^2}
(\tilde{\phi}^\dagger q^j_L). \nonumber
\end{eqnarray}
We want to express this result in the 
operator basis ${\mathcal{O}}_x$ in Table~\ref{operators}, which
requires the use of the equations of motion of $\lag_4$. We
find
\begin{eqnarray}
(\alpha^{(1)}_{\phi q})_{ij} &=&
 \frac{1}{4}V^\dagger_{ik} \lambda^{\prime *}_{k} 
\lambda^{\prime}_{l} V_{lj}, 
\nonumber \\
(\alpha^{(3)}_{\phi q})_{ij} &=&
-(\alpha^{(1)}_{\phi q})_{ij},\label{alpha:T}
\\
(\alpha_{u \phi})_{ij} &=&
\frac{1}{2} V^\dagger_{ik}\lambda^{\prime *}_{k} \lambda^\prime_j
\lambda^u_j.
\nonumber
\end{eqnarray}
The other coefficients vanish. Now we just have to
substitute these coefficients in Eq. (\ref{couplings})
to obtain
\begin{eqnarray}
X^{uL}_{ij}&=&\delta_{ij}-\frac{1}{2}\frac{v^2}{M^2} 
\lambda^{\prime *}_i \lambda^\prime_j, \nonumber \\
X^{uR}_{ij}&=&0,\quad X^{dL}_{ij}=\delta_{ij}, \quad X^{dR}_{ij}=0,
\nonumber \\
W^L_{ij}&=& 
\left(\delta_{ik}-\frac{1}{4}\frac{v^2}{M^2} \lambda^{\prime *}_i
\lambda^\prime_k \right) \tilde{V}_{kj}, \label{couplins:T} \\
W^R_{ij}&=&0,\nonumber \\
Y^u_{ij}&=&
\left(\delta_{ij}-\frac{1}{2} \frac{v^2}{M^2} (1+\frac{1}{2} \delta_{ij})
\lambda^{\prime *}_i 
\lambda^\prime_j \right) \lambda^u_j, 
\nonumber \\
Y^d_{ij}&=& \delta_{ij}\lambda^d_j, \nonumber \\
Z^{uL}_{ij}&=& 
Z^{uR}_{ij}=Z^{dL}_{ij}=Z^{dR}_{ij}=0.
\nonumber  
\end{eqnarray}
The decomposition of $W^L$ in Eq.~(\ref{WL:HLVL}) holds for
$V^L=\tilde{V}$, $H^L={\mathbf{1}}- \frac{1}{4}
\frac{v^2}{M^2}  \tilde{V}^\dagger (\lambda^{\prime *} \lambda^\prime)
\tilde{V}$ and ${H^L}^\prime={\mathbf{1}}- \frac{1}{4}
\frac{v^2}{M^2} (\lambda^{\prime *} \lambda^\prime)$, where the
initial $V$ in Eq. (\ref{full:lag:T})  is such that no further phase
redefinitions are necessary.
The coefficients in Eq.~(\ref{alpha:T}) satisfy the
relation~(\ref{prop}) with $a=-1$ and $\alpha_{\phi
q}^{(3)}$ is negative semidefinite.  
Therefore, this model fulfils
Eqs.~(\ref{XWW:prop:a}--\ref{XWd}) with $a=-1$, and
Eqs.~(\ref{X-1}--\ref{Xr2},\ref{WWgt11},\ref{WWgt1}) with ``$\leq$''.
In particular, the limits~(\ref{exp:bounds})  
and~(\ref{bounds2})
must be satisfied. Values of
the Yukawas $\lambda^\prime$ can be found such that these limits are
saturated. Basically, the maximal values are achieved when the new
quark mixes with $t$ and with either $u$ or $c$, but not with both $u$
and $c$. 
Otherwise the tight experimental bound on $X^{uL}_{uc}$ would require
a very large mass $M$, and hence a small $X^{uL}_{ut}$ and
$X^{uL}_{ct}$.  

Finally, it is interesting to see what kinds of models can produce
large new effects in quark mixing. Of course, one necessary condition
is that the high energy scale $\Lambda$ be sufficiently low ($\sim
1\mbox{TeV}$).  The other necessary condition 
(assuming weak coupling) is that the operators in $\lag^{\mathit{eff}}$
be generated at tree level. A classification of the new physics that
contributes at tree level to each process was given in
Ref.~\cite{papiro4}. 
\begin{figure}[ht]
\begin{center}
\epsfxsize=8cm
\epsfbox{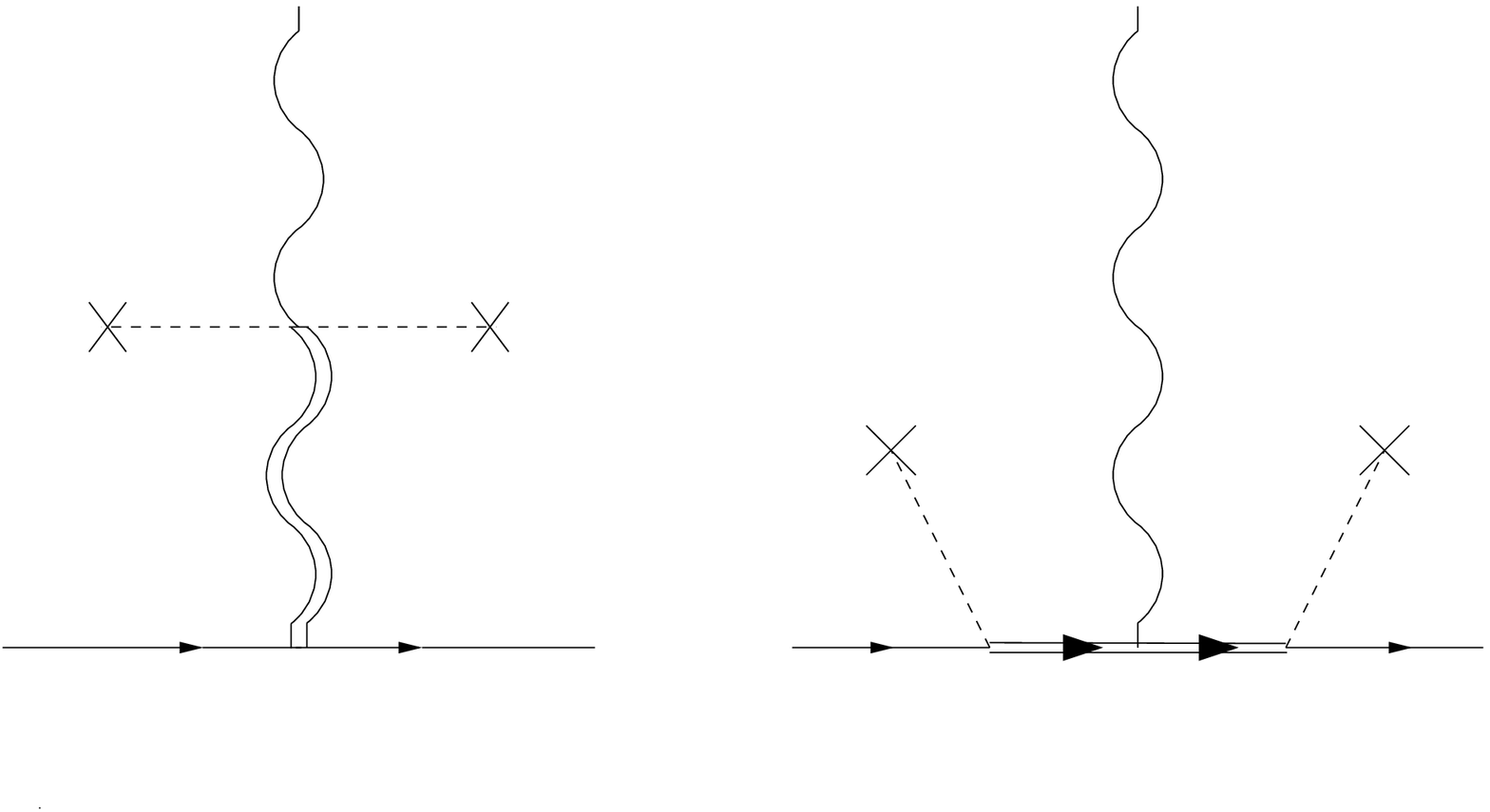}
\end{center}
\caption{Diagrams generating new $\bar{q}q^\prime V$ couplings at tree
level. The first diagram represents the exchange of a heavy vector
boson that mixes with $V$ and the second one represents the
exchange of a heavy quark that mixes with $q$ and
$q^\prime$. \label{fig}} 
\end{figure}
If the high energy theory is a renormalizable gauge theory, the only
heavy particles that generate at tree level the operators in
Table~\ref{operators} with a covariant derivative are either extra
gauge bosons or extra quarks.  
Diagrams contributing to $\bar{q}q^\prime V$ vertices are depicted in
Fig.~\ref{fig}. The new vector bosons can mix with the $Z$ and the $W$
and hence generate trilinear couplings through the first tree-level
diagram in 
Fig.~\ref{fig}~\cite{Langacker}. However, the mixing angle $\theta$
between the new and standard gauge bosons is typically
 constrained to be $\ltsim
0.01$ by the $Z$-pole data from LEP~\cite{papiro1}. Hence, besides the
suppression 
$v^2/\Lambda^2$ there is an additional suppression given by
$\theta$. Therefore the largest mixing effects arise in models with
extra quarks. 
A fourth generation of chiral fermions is excluded at the $99\%$ C.L. 
by the present limits on the S parameter~\cite{papiro1}.
Hence, the extra quarks giving rise to large mixing must be
vector-like, {\it i.e.}, their left-handed and right-handed
components must transform in the same representation of the gauge
group. These particles are present in many well-motivated 
SM extensions~\cite{papiro10}. Besides, they must couple to the
standard quarks, which 
restricts their possible representations: only singlets, doublets and
triplets under $SU(2)_L$ contribute. Their hypercharges are also
constrained. 
In Ref.~\cite{papiro9} we present a detailed study of quark mixing
in general models with exotic quarks above the electroweak
scale. Analyses of quark mixing in particular models can be found in
Ref.~\cite{papiro10}.

\section*{Acknowledgements}
It is a pleasure to thank J.A. Aguilar-Saavedra, F. Cornet,
J.L. Cort\'{e}s and J. Prades for discussions. JS thanks the
Dipartimento di Fisica ``Galileo Galilei'' and INFN sezione di Padova
for their hospitality. This work has been
supported by CICYT and Junta de
Andaluc\'{\i}a. MPV and JS also
thank MECD for financial support.


\begin{thebibliography}{99}

\bibitem{papiro6} 
S.~L.~Glashow, J.~Iliopoulos and L.~Maiani,
Phys.\ Rev.\  {\bf D2} (1970) 1285.

\bibitem{papiro5} 
N.~Cabibbo,
Phys.\ Rev.\ Lett.\  {\bf 10} (1963) 531;
M.~Kobayashi and T.~Maskawa,
Prog.\ Theor.\ Phys.\  {\bf 49} (1973) 652.


\bibitem{triangle}
L.-L. Chau and W.Y. Keung, Phys. Rev. Lett. {\bf 53}, 1802 (1984);
J.D.Bjorken, private communication and Phys. Rev. {\bf D39}, 1396
(1989); 
G.C. Branco and L. Lavoura, Phys. Lett. {\bf B208}, 123 (1988);
C. Jarlskog and R. Stora, Phys. Lett. {\bf B208}, 268 (1988);
J.L. Rosner, A.I. Sanda and M.P. Schimdt, in Proceedings of the Workshop
on High Sensitivity Beauty Physics at Fermilab, Fermilab, November 11-14,
1987, edited by A.J. Slaughter, N. Lockyer, and M. Schmidt (Fermilab,
Batavia, 1988), p. 165;
C. Harnzaoui, J.L. Rosner and A.I. Sanda, ibid., p. 215.

\bibitem{papiro7} 
M.~Beneke {\it et al.},
hep-ph/0003033.


\bibitem{papiro2} 
H.~Georgi,
``Weak Interactions And Modern Particle Theory,''
{\it Benjamin/Cummings ( 1984)};
A.~Dobado, A.~G\'{o}mez-Nicola,A.~L.~Maroto and J.~R.~Pel\'{a}ez, 
``Effective Lagrangians for the Standard Model,''
{\it  Springer (1996)}.

\bibitem{papiro3} 
W.~Buchmuller and D.~Wyler,
Nucl.\ Phys.\  {\bf B268} (1986) 621.

\bibitem{papiro4} 
C.~Arzt, M.~B.~Einhorn and J.~Wudka,
Nucl.\ Phys.\  {\bf B433} (1995) 41
[hep-ph/9405214];
C.~Arzt,
Phys.\ Lett.\  {\bf B342} (1995) 189
[hep-ph/9304230].

\bibitem{papiro8} 
T.~Appelquist, M.~J.~Bowick, E.~Cohler and A.~I.~Hauser,
Phys.\ Rev.\  {\bf D31} (1985) 1676;
E.~Bagan, D.~Espriu and J.~Manzano,
Phys.\ Rev.\  {\bf D60} (1999) 114035
[hep-ph/9809237].

\bibitem{weinberg}
J.~Gomis and S.~Weinberg,
Nucl.\ Phys.\  {\bf B469}, 473 (1996)
[hep-th/9510087].

\bibitem{papiro10}
F.~del Aguila and M.~J.~Bowick,
Nucl.\ Phys.\  {\bf B224} (1983) 107;
P.~Fishbane, S.~Meshkov and P.~Ramond,
Phys.\ Lett.\  {\bf B134} (1984) 81;
F.~del Aguila and J.~Cort\'es,
Phys.\ Lett.\  {\bf B156} (1985) 243;
G.~C.~Branco and L.~Lavoura,
Nucl.\ Phys.\  {\bf B278} (1986) 738;
J.L. Rosner, Commun. Nucl. Part. Phys. {\bf 15} (1986) 195;
V. Barger, N.G. Deshpande, R.J.N. Phillips and 
K. Whisnant, Phys. Rev. {\bf D33} (1986) 1912;
P.~Langacker and D.~London,
Phys.\ Rev.\  {\bf D38} (1988) 886;
B.~S.~Balakrishna, A.~L.~Kagan and R.~N.~Mohapatra,
Phys.\ Lett.\  {\bf B205} (1988) 345;
R.~Barbieri and L.~J.~Hall,
Nucl.\ Phys.\  {\bf B319} (1989) 1;
J.L. Hewett and T.G. Rizzo, Phys. Rep. {\bf 183}, 193 (1989);
Y. Nir and D. Silverman, Phys. Rev. {\bf D42}, 1477 (1990);
E. Nardi, E. Roulet and D. Tommasini, Phys. Rev. {\bf D46}, 3040 (1992);
V. Barger, M.S. Berger and R.J.N. Phillips, Phys. Rev. {\bf D52}, 1663
(1995) [hep-ph/9503204];
P.~H.~Frampton, P.~Q.~Hung and M.~Sher,
Phys.\ Rep.\  {\bf 330}, 263 (2000)
[hep-ph/9903387].

\bibitem{papiro9} 
F.~del Aguila, M. P\'{e}rez-Victoria and J. Santiago, UGFT-118/00,
MIT-CTP-2997, hep-ph/0007316.

\bibitem{papiro11} 
F.~del Aguila, J.~A.~Aguilar-Saavedra and R.~Miquel,
Phys.\ Rev.\ Lett.\  {\bf 82} (1999) 1628
[hep-ph/9808400].

\bibitem{BSW}
S.~Bar-Shalom and J.~Wudka,
Phys.\ Rev.\  {\bf D60}, 094016 (1999)
[hep-ph/9905407].

\bibitem{papiro12}
L.~Wolfenstein,
Phys.\ Rev.\ Lett.\  {\bf 51} (1983) 1945.

\bibitem{papiro1} 
D.~E.~Groom {\it et al.},
Eur.\ Phys.\ J.\  {\bf C15} (2000) 1;
M.~Martinez, R.~Miquel, L.~Rolandi and R.~Tenchini,
Rev.\ Mod.\ Phys.\  {\bf 71} (1999) 575.

\bibitem{Langacker}
P.~Langacker and M.~Plumacher,
Phys.\ Rev.\  {\bf D62}, 013006 (2000)
[hep-ph/0001204].




\end{thebibliography}
\end{document}